\documentclass[aps,pra,twocolumn,superscriptaddress,showpacs,showkeys,amsmath,amssymb]{revtex4}
\usepackage{amsfonts}
\usepackage{amssymb,amsmath}
\usepackage{mathrsfs}
\usepackage{latexsym}
\usepackage{amsmath}
\usepackage[cp1251]{inputenc}
\usepackage{graphicx}
\usepackage{dcolumn}
\usepackage{bm}
\usepackage{color}

\RequirePackage{ifthen}
\RequirePackage[pdfstartview=FitH]{hyperref}
\begin{document}
	
\title{Four-body physics in low-dimensional bosons with three-body interaction}
\author{V.~Polkanov\footnote{e-mail: cogersum92@gmail.com}}	
\author{O.~Hryhorchak\footnote{e-mail: hrorest@gmail.com}}
\author{V.~Pastukhov\footnote{e-mail: volodyapastukhov@gmail.com}}
\affiliation{Professor Ivan Vakarchuk Department for Theoretical Physics, Ivan Franko National University of Lviv, 12 Drahomanov Street, Lviv, Ukraine}

	\date{\today}

	
	
\begin{abstract}
The two-channel model for bosons with the three-body interaction is proposed. Similar to the Hamiltonian describing narrow Feshbach resonance in the two-body sector, our model includes the finite-range effects of the three-body potential and is well-defined in the ultraviolet (UV). A detailed exploration of the Efimov-like effect in the fractal-dimension system of four bosons is carried out. Peculiarities of the four-body bound states and the low-energy atom-trimer scattering in one dimension are revealed.
\end{abstract}
	
	\maketitle
\section{Introduction}
In contrast to the condensed-matter systems, where collective effects are determinative in forming their properties, the behavior of the ultracold gases is typically strongly influenced by the few-body physics. When details of the interparticle interaction potential can be neglected, the latter becomes universal, governing \cite{Tan,Braaten_2008} short-distance (large-momentum) behavior of various observables. On the other hand, by tuning parameters of the microscopic interaction to the resonant (or scaling) limit, one can expect the emergence of highly nontrivial, essentially quantum effects in systems with few additional particles.

One of the most famous examples is the Efimov effect \cite{Efimov_70, Naidon_Endo_17, Greene_et_al} which consists of emerging the infinite tower of energy levels ($\epsilon_{3, n} \sim e^{\frac{-2\pi n}{s_0}}$, for large $n$, where $s_0$ is constant) of the three-body sector of three-dimensional bosonic system with the two-body interaction in scaling limit. Due to strong long-ranged forces induced \cite{Efimov_73} at significant atomic mass imbalance, a similar behavior was observed \cite{Petrov_03} for three-dimensional two-component fermions in the $p$-wave channel of their three-body sector. Along with that, in two dimensions, the super Efimov effect between identical spin-polarized fermions was discovered \cite{Nishida_13}, consisting on emerging an infinite number of binding states with energy levels reducing to zero in double exponent law [$\epsilon_{3, n} \sim \exp(-2 e^{\frac{3\pi n}{4}}+\theta)$ for large $n$, where $\theta$ is nonuniversal constant]. Later on, the semisuper Efimov effect ($\epsilon_{4, n} \sim e^{-\frac{(\pi n)^2}{27}}$, for large $n$) was predicted \cite{Nishida_17} in four-particle system of two-dimensional bosons with the short-ranged three-body interaction. A purely Efimov's behavior of the energy levels for four spineless bosons or three-component fermions with the suppressed two-body interaction can be found \cite{Hryhorchak_22,Polkanov_24} only in fractional dimensions.

The approximate nature of the through-field interaction description \cite{Primakoff_1939} creates a series of interparticle potentials: two-body, three-body, etc. Interplay of short-range repulsive part and attractive tail of realistic two-body potential together with quantum effects produces effective three- \cite{Valiente_19,Pricoupenko_Petrov_19} and higher-order \cite{Pricoupenko_Petrov_21} interactions. A strong external potential restricting the system in one or two directions induces \cite{Mazets_08} effective $N$-body interaction in unconfined dimensions. Recently, the emergence of the three-body attraction in the two-component system of interacting, via two-body potentials, bosons with Rabi coupling was demonstrated \cite{Hammond_22,Tiengo_25}. Although in most cases an interatomic interaction is well described by two-body potentials, the three-body interaction, because of its significant impact on low-dimensional quantum systems, cannot be ignored when studying their few-body properties and collective behavior. Moreover, as in the case of the Feshbach resonance \cite{Feshbach_1962}, the two-body interaction can be completely switched off, allowing the direct investigation of the three-body interaction effects. 

In the last decade, systems with a local three-body potential that demonstrates the intrinsic scale anomaly \cite{Drut_18,Daza_19} in one dimension, were extensively studied in few-body \cite{Nishida_18,Guijarro_et_al,McKenney_19,Liu_21} and many-body \cite{Sekino_18,Pastukhov_19,Valiente_Pastukhov,Maki_19,Czejdo_20,McKenney_20,Morera_22} setups. A characteristic feature of the three-body pseudopotential \cite{Pricoupenko_19} is the existence of a vacuum trimer, while at finite densities, another exotic tripling mechanism can be realized \cite{Akagami_21,Tajima_22}. In macroscopic systems, therefore, there is always a competition between trimer dissociation and dimer formation \cite{Hryhorchak_24}, and the coexistence of trimers and unbound atoms in the strong coupling regime qualitatively discussed in Ref.~\cite{Tajima_24}. To address this problem quantitatively, one needs to consider the impact of atom-trimer, dimer-trimer, and trimer-trimer collisions. Part of the present work deals with the solution of the former problem, at least for bosons interacting through the contact three-body potential. By utilizing the two-channel model that incorporates finite-range effects of the three-body interaction, we also present a detailed study of the Efimov-like behavior of the four-body sector in fractional dimensions.

\section{Model}
The considered model describes spinless particles (of mass $m$) loaded in the low-dimensional geometries with fractional $d$s (with $d=1$ included). The two-body scattering length is assumed to be tuned to the resonant magnitude, i.e., the two-body contact interaction vanishes. From the effective field theory perspectives, below $d=2$ local gradient terms (of the form $\psi^{\dagger}\psi\nabla^2 \psi^{\dagger}\psi$) that take into account the finite-range corrections to the two-body potential, are less relevant than the direct three-body contact interaction. This suggests that the low-energy physics of the low-dimensional bosons with divergent two-body scattering length is controlled by local three-body interaction term $\frac{1}{3!}g_{3,\Lambda}(\psi^{\dagger})^3\psi^3$ [where $\Lambda$ is UV cutoff]. This theory is renormalizable only in 1D, and from previous analysis \cite{Hryhorchak_22}, we already know that it is UV incomplete in four and higher order particle sectors in $d>1$ due to the emergence of the Efimov-like effect. To cure this issue, one needs to introduce explicitly into the system, except for the three-body scattering length (or width of an appropriate bound state), one more parameter with dimension of length. This can be most simply realized by considering the two-channel model (similar to that introduced recently for fermions \cite{Polkanov_24}) with the Hamiltonian
\begin{eqnarray}\label{H}
&&H=\int_{\bf p}\varepsilon_{\bf p}\psi^{\dagger}_{\bf p}\psi_{\bf p}+\int_{\bf p}\left\{\frac{\varepsilon_{\bf p}}{3}+\delta\omega_{\Lambda}\right\}\Psi^{\dagger}_{\bf p}\Psi_{\bf p}\nonumber\\
&&+\frac{g}{3!} \int_{{\bf p}_1,{\bf p}_2,{\bf p}_3}\{\Psi^{\dagger}_{{\bf p}_1+{\bf p}_2+{\bf p}_3}\psi_{{\bf p}_1}\psi_{{\bf p}_2}\psi_{{\bf p}_3}+\textrm{h.c.}\},
\end{eqnarray}
where integration $\int_{\bf p}\equiv\int\frac{d{\bf p}}{(2\pi)^d}$ is carried out over the $d$-dimensional wave-vector inside sphere of large radius $\Lambda$. Bosonic creation (annihilation) operators $\psi^{\dagger}_{\bf p}$ ($\psi_{\bf p}$) and $\Psi^{\dagger}_{\bf p}$ ($\Psi_{\bf p}$) describe bosons and the closed-channel molecules (trimers), respectively, satisfy the canonical commutation relations $[\psi_{\bf p}, \psi^{\dagger}_{{\bf p}'}]=[\Psi_{\bf p}, \Psi^{\dagger}_{{\bf p}'}]=(2\pi)^d\delta({\bf p}-{\bf p}')$ (with all other pairs commuting). We have also introduced shorthand notations for the Galilean-invariant free-atom dispersion $\varepsilon_{\bf p}=\frac{{\bf p}^2}{2m}$ and for the cutoff-dependent detuning parameter $\delta\omega_{\Lambda}=-g^2/(3!g_{3,\Lambda})$. The inter-channel coupling constant $g=1/(mr^{2-d}_0)$ depends on the parameter $r_0$ (with dimension of length) explicitly related to the effective range of microscopic three-body potential. Another parameter with dimension of length in Hamiltonian (\ref{H}) is determined by the three-body problem. It can be related \cite{Hryhorchak_22,Polkanov_24} to the broad-resonance ($g\to \infty$) bound state energy of three bosons $|\epsilon_{\infty}|$. In this limit model (\ref{H}) is equivalent to a system of bosons (without closed-channel molecules) interacting through the contact three-body pseudo-potential with bare coupling constant
\begin{eqnarray}\label{g_3}
g^{-1}_{3,\Lambda}=-\int_{{\bf p}, {\bf p}'}\frac{1}{\varepsilon_{{\bf p}}+\varepsilon_{{\bf p}'}+\varepsilon_{{\bf p}+{\bf p}'}+|\epsilon_{\infty}|}.
\end{eqnarray}
Analogously to the system of fermions \cite{Polkanov_24}, we can solve the three-body problem for bosons at finite magnitudes of the effective range parameter (finite $g$s). Skipping the details of these simple calculations, we only present a transcendental equation on the three-body bound state energy
\begin{eqnarray}\label{e_g}
\epsilon_{g}+\frac{g^2}{3!g_3}\left[1-\left(\frac{\epsilon_g}{\epsilon_{\infty}}\right)^{d-1}\right]=0,
\end{eqnarray}
(note $\epsilon_{g\to \infty}\to\epsilon_{\infty}$) where the `observable' coupling constant reads
\begin{eqnarray}
g^{-1}_3=-\frac{\Gamma(1-d)}{(2\sqrt{3}\pi)^d}m^d|\epsilon_{\infty}|^{d-1}.
\end{eqnarray}
In contrast to $g_{3,\Lambda}$, this parameter determines all observable properties of the system at broad resonance, for instance, the three-body scattering amplitude. Therefore, if one puts three bosons in a large box of volume $L^d$ with the periodic boundary conditions, the energy of the lowest scattering state is asymptotically shifted by $\frac{g_3}{3!L^d}$ in the $L\to \infty$ limit. Note that the unitary limit is reached for zero and finite effective ranges when $g^{-1}_3=0$. Within definition (\ref{g_3}), the three-body bound-state energy relation (\ref{e_g}) is well-defined in one dimension
\begin{eqnarray}\label{e_g_d_1}
\epsilon_{g}=\frac{g^2m}{12\sqrt{3}\pi}\ln\left(\frac{\epsilon_g}{\epsilon_{\infty}}\right).
\end{eqnarray}
Finally, it should be stressed that the effective model described by the Hamiltonian (\ref{H}) (at finite $g$s) possesses not only the UV-complete three-body sector but also any few-body physics of bosons with the three-body interaction is well-defined even in $d>1$.

\section{Four-body problem}
Any four-body state (bound or scattering) of the considered model Hamiltonian can be written down in the following form (in the center-of-mass frame):
\begin{align}\label{4_state}
&|4\rangle=\int_{{\bf p}}C_{{\bf p}}\Psi^{\dagger}_{-{\bf p}}\psi^{\dagger}_{{\bf p}}|0\rangle\nonumber\\
&+\int_{\sum_i{\bf p}_{i}=0}C_{{\bf p}_{1},{\bf p}_2,{\bf p}_{3},{\bf p}_{4}}
\psi^{\dagger}_{{\bf p}_1}\psi^{\dagger}_{{\bf p}_2}\psi^{\dagger}_{{\bf p}_3}\psi^{\dagger}_{{\bf p}_{4}}|0\rangle,
\end{align}
where $|0\rangle$ is the normalized vacuum state, and amplitudes $C_{{\bf p}}$,  $C_{{\bf p}_{1},{\bf p}_2,{\bf p}_{3},{\bf p}_{4}}$ are subject to the Schr\"odinger equation. The bosonic statistics requires $C_{{\bf p}_{1},{\bf p}_2,{\bf p}_{3},{\bf p}_{4}}$ to be a fully symmetric function of its four arguments. Substituting wave-function into equation $H|4\rangle=\mathcal{E}|4\rangle$ with $\mathcal{E}$ being the eigenvalue of the Hamiltonian (\ref{H}), we obtain the system of two coupled equations. The amplitude $C_{{\bf p}_{1},{\bf p}_2,{\bf p}_{3},{\bf p}_{4}}$ is determined by simple algebraic equation with a solution
\begin{eqnarray}\label{C_pppp}
C_{{\bf p}_{1},{\bf p}_2,{\bf p}_{3},{\bf p}_{4}}=-\frac{g}{4!}\frac{\sum_i C_{{\bf p}_i}}{\sum_i \varepsilon_{{\bf p}_i}-\mathcal{E}},
\end{eqnarray}
for negative $\mathcal{E}$s. The function $C_{{\bf p}}$ is the solution to integral equation
\begin{eqnarray}\label{C_p_Eq}
\mathcal{D}^{-1}_{{\bf p}}(\mathcal{E})C_{{\bf p}}=\frac{g^2}{2}\int_{{\bf p}'}\Pi_{{\bf p},{\bf p}'}(\mathcal{E})C_{{\bf p}'},
\end{eqnarray}
where $\mathcal{D}^{-1}_{{\bf p}}(\mathcal{E})=\frac{4}{3}\varepsilon_{{\bf p}}-\mathcal{E}-\frac{g^2}{3!}t^{-1}_{{\bf p}}(\mathcal{E})$, $t_{{\bf p}}(\mathcal{E})=t_3({\bf p}, \varepsilon_{{\bf p}}-\mathcal{E})$ with the three-body $t$-matrix defined as follows:
\begin{eqnarray}\label{t_3}
t^{-1}_3({\bf p}, \omega)=g^{-1}_{3}\left[1-\frac{(\varepsilon_{\bf p}/3-\omega)^{d-1}}{|\epsilon_\infty|^{d-1}}\right].
\end{eqnarray}
The kernel of the integral equation (\ref{C_p_Eq}) is related to the particle-particle scattering bubble diagram and can be explicitly computed
\begin{eqnarray}
&&\Pi_{{\bf p},{\bf p}'}(\mathcal{E})=\frac{\Gamma\left(1-d/2\right)}{(4\pi)^{d/2}}m^{d/2}\nonumber\\
&&\times\left(\frac{1}{2}\varepsilon_{{\bf p}+{\bf p}'}+\varepsilon_{\bf p}+\varepsilon_{{\bf p}'}-\mathcal{E}\right)^{d/2-1}.
\end{eqnarray}
It is readily seen that the above equations describing four-body states with {\it negative} energies are valid in one dimension. In particular, taking into account the definition (\ref{g_3}) of the `observable' coupling $g_3$, one recovers from Eq.~(\ref{t_3}) the correct expression \cite{Valiente_Pastukhov,Valiente_19} for the three-body $t$-matrix in the $d\to 1$ limit. A further discussion will be fully devoted to finding solutions to Eq.~(\ref{C_p_Eq}). However, their construction is totally different for bound and scattering states and should be considered separately.

\subsection{Bound states}
\begin{figure}[h!]
	\centerline{\includegraphics
		[width=0.45
		\textwidth,clip,angle=-0]{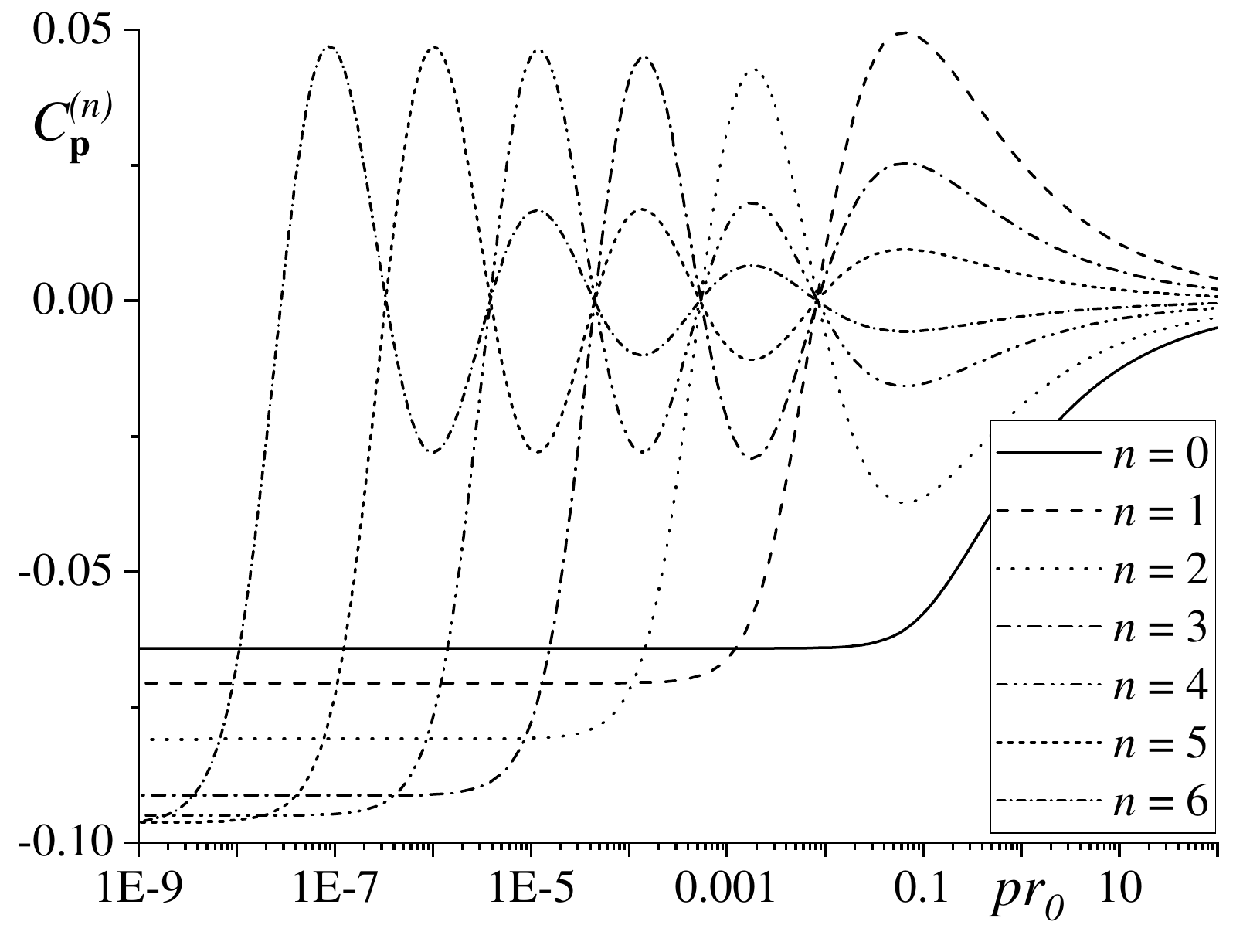}}
	\caption{Spherically-symmetric unnormalized wavefunctions corresponding to different universal (Efimov-like) four-body bound states $\epsilon_{4,n}$.}\label{C_p_Efimov_fig}
\end{figure}

Denoting the four-body bound state energy by $\epsilon_4$, one should be aware of the fact that $|\epsilon_4|>|\epsilon_g|$ (otherwise the tetramer disappears in the trimer-atom continuum). In arbitrary dimension $d$, the total angular momentum of the four-body system is conserved likewise parity in 1D, which should be reflected in the solutions $C_{\bf p}$. Therefore, one can expand both sides of Eq.~(\ref{C_p_Eq}) in series over the Gegenbauer polynomials \cite{Abramowitz} sequentially equating coefficients. The $s$-wave bound states are energetically most favorable to emerge; therefore, we focus exclusively on them (and even wave functions in 1D) below. Mathematically, the bound-state problem reduces to the search for eigenvalues and eigenfunctions of the appropriate linear integral operator generating Eq.~(\ref{C_p_Eq}). In Ref.~\cite{Hryhorchak_22}, it was demonstrated, utilizing a model with the three-body contact potential, that the four-boson sector possesses the Efimov-like effect with an infinite tower of bound states with the universal ratio $\epsilon_{4,n+1}/\epsilon_{4,n}=e^{-2\pi/\eta}$ of the energy levels at large $n$. The lowest bound state is inaccessible in the broad-resonance model (it does not exist at all) due to Thomas-like collapse. At the same time, exponent $\eta$ is a pure number that depends only on the spatial dimensionality. Hamiltonian (\ref{H}) at finite effective ranges instead is self-adjoint and allows the calculations of the full spectrum. In Table.~\ref{tab:table1}
\begin{table}[h!]
	\begin{center}
		\begin{tabular}{c|c|c} % {l|c|r} <-- Alignments: 1st column left, 2nd middle and 3rd right, with vertical lines in between
			%			\textbf{Value 1} & \textbf{Value 2} & \textbf{Value 3}\\
			$n$ & $\kappa_n$ & $\kappa_{n-1}/\kappa_{n}$ \\
			\hline
			0  & 0.0680813794943696    & \\
			1  & 0.0024777093701106  & 27.47755\\
			2  & 0.0002000665387490   & 12.38443\\
			3  & 1.7124230575426005$\times10^{-5}$  & 11.68324\\
			4  & 1.4726731645048760$\times10^{-6}$  & 11.62799\\			
			5  & 1.2671106854818457$\times10^{-7}$  & 11.62229\\
			6  & 1.0903192403477262$\times10^{-8}$  & 11.62146\\
			7  & 9.382509949064195$\times10^{-10}$  & 11.62076\\
			$\vdots$ & $\vdots$               & $\vdots$\\
			$\infty$ & 0                      & 11.5941\\
		\end{tabular}
		\caption{Dimensionless parameter determining the first few four-body bound states in dimension $d=1.59$ at unitarity. }\label{tab:table1}
	\end{center}
\end{table}
we presented first few energy levels (parameterized by $\epsilon_{4,n}=-\frac{2\kappa^2_n}{3mr^2_0}$) for dimension $d=1.59$, where the characteristic universal scaling factor $e^{\pi/\eta}=11.5941\dots$ is close to minimal value. We considered unitary limit $\epsilon_g=0$ ($g_3=\infty$), which is the most convenient for observing the Efimov effect. The results demonstrate a very slow convergence of the scaling factor to its universal value. This contrasts \cite{Griesshammer_23} to the three-dimensional bosons with pairwise contact interaction. Even for low-dimensional fermions in the four-particle $p$-wave states \cite{Polkanov_24}, a better-behaved kernel of the integral equation provides faster convergence. The numerical diagonalization of the integral operator in Eq.~(\ref{C_p_Eq}) also yields the appropriate eigenfunctions $C^{(n)}_{{\bf p}}$ (see Fig.~\ref{C_p_Efimov_fig}).
These curves confirm the logarithmically periodic behavior of the wave function in the scaling limit $|{\bf p}|\ll 1/r_0$.

\begin{figure}[h!]
	\centerline{\includegraphics
		[width=0.45
		\textwidth,clip,angle=-0]{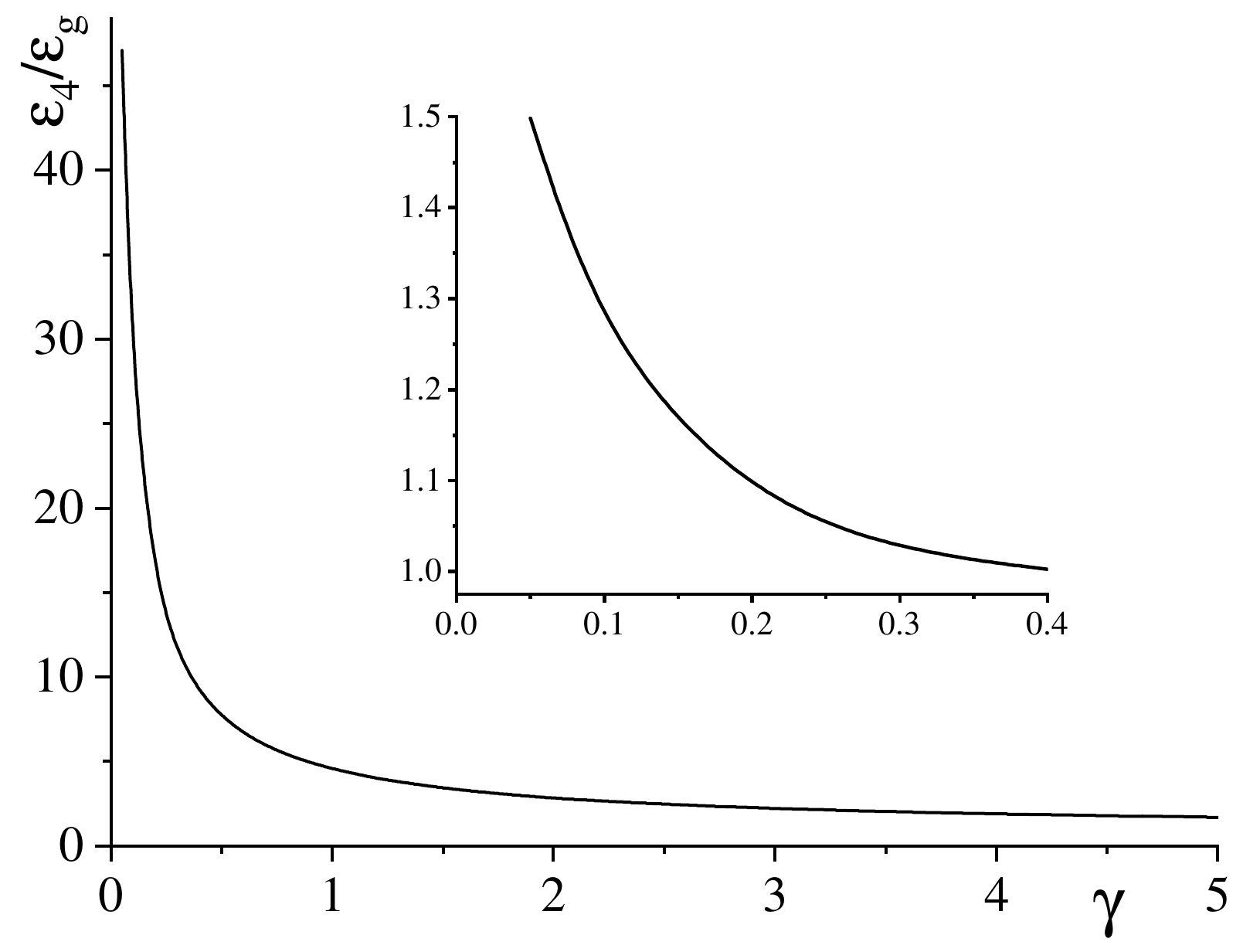}}
	\caption{Four-body bound states $\epsilon_{4}$ in one dimension. The first excited state occurs at $\gamma\approx 0.4$ (see insert).}\label{bind_en_1d_fig}
\end{figure}

Although there is no Efimov-like effect in our system in 1D, the bound states exist. Even though the three-body contact interaction is renormalizable in 1D, the numerical solution of the four-body problem at narrow resonance requires massive computational efforts. We have analyzed the effect of non-zero effective ranges on the four-body problem. It is natural to measure the distance to broad resonance using the deviation of parameter $\gamma=\ln\left(\frac{\epsilon_{\infty}}{\epsilon_{g}}\right)$ from zero. The results for the dimensionless ratio $\epsilon_{4}/\epsilon_{g}$ are plotted in Fig.~\ref{bind_en_1d_fig}. 
In particular, they demonstrate a rapid increase of $\epsilon_{4}$ at small $\gamma$ and reveal the emergence (see insert) of the first excited four-body bound state at $\gamma\approx 0.4$. This second bound state is crucial for correctly interpreting the four-body scattering solutions.

\subsection{Atom-trimer scattering states}
\begin{figure}[h!]
	\centerline{\includegraphics
		[width=0.45
		\textwidth,clip,angle=-0]{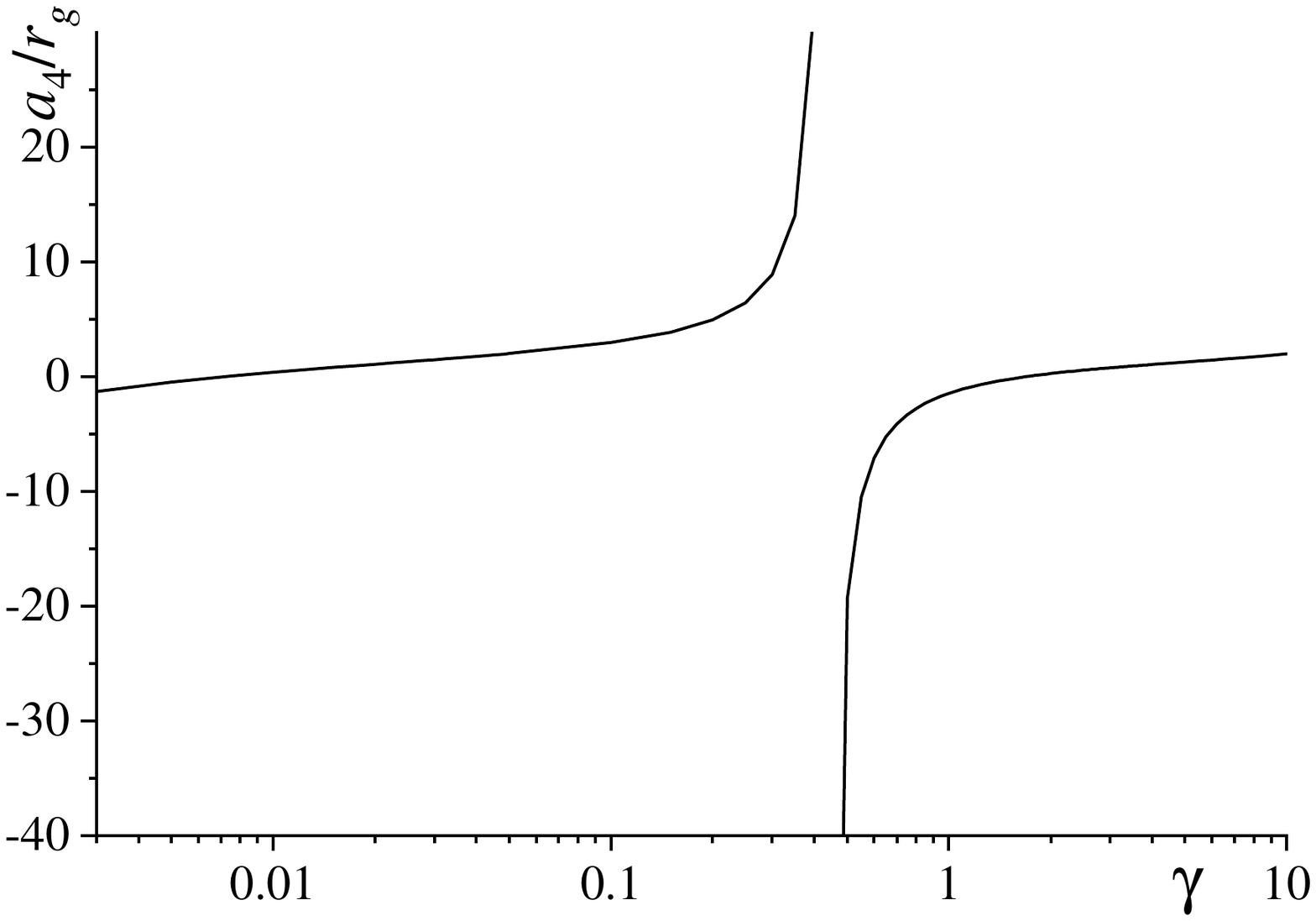}}
	\caption{Atom-trimer scattering length $a_4$ (in units of $r_g$, where $\epsilon_g=-\frac{2}{3mr^2_g}$) as function of $\gamma=\ln\left(\frac{\epsilon_{\infty}}{\epsilon_{g}}\right)$.}\label{1d_scatt_fig}
\end{figure}

Let us consider a special class of negative-energy four-body states involving the elastic scattering of a single boson on the trimer. For zero total momentum, the energy of the system is fixed as $\mathcal{E}_{\bf k}=\epsilon_g+\frac{4}{3}\varepsilon_{\bf k}$, where the last term is called collision energy and refers to the total kinetic energy of trimer and atom moving with opposite momenta ${\bf k}$ and $-{\bf k}$. Since $\mathcal{D}_{{\bf p}}(\mathcal{E}_{\bf k})$ has simple poles at $|{\bf p}|=|{\bf k}|$, the solution to Eq.~(\ref{C_p_Eq}) in this case should reflect the physics of the atom-trimer scattering. Particularly, the partial solution to a non-homogeneous equation must be chosen in a way that reduces to the outgoing spherical wave at infinity. Technically it is done by shifting a pole from the real axis through the `$i0_+$-prescription' $\mathcal{E}_{\bf k}\to \mathcal{E}_{\bf k}+i0_+$ (with $0_+$ being infinitesimal positive)
\begin{align}\label{C_p_scatt_Eq}
C_{{\bf p}}=C^{(0)}_{{\bf p}}+\frac{g^2}{2}\mathcal{D}_{{\bf p}}( \mathcal{E}_{\bf k}+i0_+)\int_{{\bf p}'}\Pi_{{\bf p},{\bf p}'}(\mathcal{E}_{\bf k})C_{{\bf p}'},
\end{align}
where in the last term, $i0_+$ has no effect if the collision energy is below the threshold, i.e., $\mathcal{E}_{\bf k}<0$. Throughout numerous solutions $C^{(0)}_{{\bf p}}$ to the homogeneous equation, the one satisfying the boundary conditions should be written down in Eq.~(\ref{C_p_scatt_Eq}). Let us first consider the zero-range (formally $g\to \infty$) model with contact three-body interaction to reveal them. Without closed-channel molecules only the second term in Eq.~(\ref{4_state}) survives with the solution (\ref{C_pppp}), and modified Eq.~(\ref{C_p_scatt_Eq}) [the $g\to \infty$ limit suggests replacement $\frac{g^2}{2}\mathcal{D}_{{\bf p}}(\mathcal{E}_{\bf k}+i0_+)\to -3t_{{\bf p}}(\mathcal{E}_{\bf k}+i0_+)$]. Necessary boundary conditions are most simply formulated in coordinate representation where the wavefunction of the four-body system reads $C({\bf r}_1,{\bf r}_2,{\bf r}_3,{\bf r}_4)\propto \int_{\sum_i{\bf p}_{i}=0}e^{i\sum_i{\bf p}_{j}{\bf r}_{j}}C_{{\bf p}_{1},{\bf p}_2,{\bf p}_{3},{\bf p}_{4}}$. The specifics of the considered system assumes the following behavior if one atom (say, with coordinate ${\bf r}_1$) is placed far away $\left|{\bf r}_1-\frac{{\bf r}_2+{\bf r}_3+{\bf r}_4}{3}\right|\to \infty$ from another three:
\begin{eqnarray}
C({\bf r}_1,{\bf r}_2,{\bf r}_3,{\bf r}_4)\to e^{i{\bf k}\left({\bf r}_1-\frac{{\bf r}_2+{\bf r}_3+{\bf r}_4}{3}\right)}\phi({\bf r}_2,{\bf r}_3,{\bf r}_4),
\end{eqnarray}
where $\phi({\bf r}_2,{\bf r}_3,{\bf r}_4)$ is the three-body bound-state wavefunction. Substituting solution (\ref{C_pppp}) in definition of $C({\bf r}_1,{\bf r}_2,{\bf r}_3,{\bf r}_4)$, and making use of the above boundary condition, one shows that $C^{(0)}_{{\bf p}}=(2\pi)^d\delta({\bf p}-{\bf k})$. Recall, such a condition we have obtained assuming zero-range limit ($g\to \infty$). However, in the opposite $g\to 0$ limit, the atom is uncoupled to the closed-channel molecule, and consequently, the same relation holds. Therefore, statement that $C^{(0)}_{{\bf p}}\propto \delta({\bf p}-{\bf k})$ is true for any magnitude of the effective range. Physically, this behavior provides the incident wave in the scattering solution. Conventionally, a solution to Eq.~(\ref{C_p_scatt_Eq}) is written down in following form:
\begin{align}\label{}
C_{{\bf p}}=(2\pi)^d\delta({\bf p}-{\bf k})+\frac{f_{{\bf p}}(\mathcal{E}_{\bf k})}{{\bf p}^2-{\bf k}^2-i0_+},
\end{align}
where the introduced function $f_{{\bf p}}(\mathcal{E}_{\bf k})$ being put on-shell $|{\bf p}|=|{\bf k}|$, reproduces the atom-trimer scattering amplitude. The off-shell function $f_{{\bf p}}(\mathcal{E}_{\bf k})$ is determined by the linear integral equation
\begin{align}\label{f_pE}
f_{{\bf p}}(\mathcal{E}_{\bf k})=h_{{\bf p}}(\mathcal{E}_{\bf k})\left\{\pi_{{\bf p},{\bf k}}(\mathcal{E}_{\bf k})+\int_{{\bf p}'}\frac{\pi_{{\bf p},{\bf p}'}(\mathcal{E}_{\bf k})f_{{\bf p}'}(\mathcal{E}_{\bf k})}{{{\bf p}'}^2-{\bf k}^2-i0_+}\right\},
\end{align}
with notations for functions $h_{{\bf p}}(\mathcal{E}_{\bf k})=({\bf p}^2-{\bf k}^2)\mathcal{D}_{{\bf p}}(\mathcal{E}_{\bf k})Z^{-1}$ and $\pi_{{\bf p},{\bf p}'}(\mathcal{E}_{\bf k})=\frac{g^2Z}{2}\Pi_{{\bf p},{\bf p}'}(\mathcal{E}_{\bf k})$, where $Z$ is the residue of $\mathcal{D}_{{\bf p}}(\mathcal{E}_{\bf k}+i0_+)$ in the pole. The structure of Eq.~(\ref{f_pE}), which is very similar \cite{Levinsen_2011} to the atom-dimer scattering problem for the closed-channel Hamiltonian involving two-body sector, allows the partial-wave decomposition in $d>1$ due to the angular momentum conservation of four particles. In the following, however, we only focus on the one-dimensional case. At very low colliding momenta, when the de Broglie wavelengths associated with atom and trimer are the largest parameters with dimension of length, the on-shell scattering amplitude $f_{{\bf k}}(\mathcal{E}_{\bf k})$ should reproduce the one for two point-like particles (of masses $m$ and $3m$) with contact two-body interaction. This observation allows us to identify the atom-trimer $s$-wave scattering length $a_4$. The appropriate effective coupling constant conventionally is written down as follows: $g_{4}=-\frac{4}{3ma_4}$. By matching the on-shell solution to Eq.~(\ref{f_pE}) at small $k$ to the symmetric part of scattering amplitude for a $\delta$-potential, $f_{\bf k}(\mathcal{E}_{\bf k})|_{k\to 0}=\frac{2ik}{1+ika_4}$, we estimated in Fig.~\ref{1d_scatt_fig}
the scattering length $a_4$ as function of parameter $\gamma$ in 1D. In our numerical calculations the length scale is set by $r_g$, which parametrizes $\epsilon_g=-2/(3mr^2_g)$ the width of the three-body bound state at finite ranges (the relation to original range is the following: $\frac{r^2_0}{r^2_g}=\frac{\ln \gamma}{8\sqrt{3}\pi}$). The results reveal the nature of the effective boson-trimer interaction, at least in relation to scattering properties in 1D: the coupling $g_4$ is mainly attractive ($g_4<0$), but there are two regions at small $\gamma<0.007$ and in interval $\gamma\in [0.43, 1.73]$, where the repulsion realizes. These circumstances are very important for the stability of the system in the thermodynamic limit. Expectedly, at $\gamma\approx 0.4$, the boson-trimer scattering is resonant due to the emergence of the excited four-body bound state. By the computations of Sekino and Nishida \cite{Sekino_18}, the three four-body bound states should exist in a broad resonance. Therefore, at smaller $\gamma$s -- the region that is not numerically accessible for us -- one more resonance should occur.

\section{Summary}
In conclusion, we have proposed the two-channel model for spinless particles with the three-body interaction and analyzed its few-body properties in lower dimensions. Due to the finite effective range, the theory is well-defined at UV in any $N$-body sector. Therefore, after renormalization of the three-body problem, we were able to study in detail the emergence of the Efimov-like effect in the four-body system in fractional $d$s and confirm the universal scaling of the appropriate energy levels. The calculation of the ground and first seven excited bound states for an arbitrary dimension $d=1.59$ revealed a slow convergence of the scaling factor to its universal value. The effective-range dependence of the four-body bound and scattering states was calculated in a one-dimensional case. For large effective ranges, the four-body system possesses a single bound state. Closer to broad resonance, the first excited bound states emerge, producing the atom-trimer $s$-wave scattering length divergence.

\begin{center}
	{\bf Acknowledgements}
\end{center}
This work was partly supported by Project No.~0122U001514 from the Ministry of Education and Science of Ukraine.

\end{document}